\documentclass[12pt,preprint]{aastex}
\usepackage{amssymb,amsmath}
\usepackage{color,hyperref}
\definecolor{linkcolor}{rgb}{0,0,0.25}
\hypersetup{
  colorlinks=true,        
  linkcolor=linkcolor,    
  citecolor=linkcolor,    
  filecolor=linkcolor,    
  urlcolor=linkcolor      
}
\newcommand{\etal}{et al.}
\newcommand{\eg}{e.g.}
\newcommand{\eqnname}{equation}

\newcommand{\sectionname}{$\mathsection$}

\newcommand{\dd}{\mathrm{d}}
\newcommand{\Gaia}{\emph{Gaia}}
\newcommand{\apogee}{APOGEE}
\newcommand{\hermes}{HERMES}
\newcommand{\hipparcos}{\emph{Hipparcos}}
\newcommand{\vR}{\ensuremath{v_R}}
\newcommand{\vphi}{\ensuremath{v_{\phi}}}
\newcommand{\vo}{\ensuremath{v_0}}
\newcommand{\Ro}{\ensuremath{R_0}}

\newcommand{\Ab}{\ensuremath{A_b}}
\newcommand{\Rb}{\ensuremath{R_b}}
\newcommand{\Omegab}{\ensuremath{\Omega_b}}
\newcommand{\fdehnen}{\ensuremath{f_{\text{Dehnen}}}}
\newcommand{\sigmaR}{\ensuremath{\sigma_R}}
\newcommand{\rE}{\ensuremath{R_e}}
\newcommand{\Lc}{\ensuremath{L_c}}
\newcommand{\Rs}{\ensuremath{R_s}}
\newcommand{\Rsigma}{\ensuremath{R_{\sigma}}}
\newcommand{\Rolr}{\ensuremath{R_{\text{OLR}}}}

\begin{document}

\title{Tracing the Hercules stream around the Galaxy}
\author{Jo~Bovy}
\affil{Center for Cosmology and Particle Physics, Department of Physics, New York University, 4 Washington Place, New York, NY 10003, USA}
\email{jb2777@nyu.edu}

\begin{abstract}
It has been proposed that the Hercules stream, a group of co-moving
stars in the Solar neighborhood offset from the bulk of the velocity
distribution, is the result of resonant interactions between stars in
the outer disk and the Galactic bar. So far it has only been seen in
the immediate Solar neighborhood, but the resonance model makes a
prediction over a large fraction of the Galactic disk. I predict the
distribution of stellar velocities and the changing Hercules feature
in this distribution as a function of location in the Galactic disk in
a simple model for the Galaxy and the bar that produces the observed
Hercules stream. The Hercules feature is expected to be strong enough
to be unambiguously detected in the distribution of line-of-sight
velocities in selected directions. I identify quantitatively the most
promising lines of sight for detection in line-of-sight velocities
using the Kullback-Leibler divergence between the predictions of the
resonance model and an axisymmetric model; these directions are at
$250^{\circ} \lesssim l \lesssim 290^{\circ}$. The predictions
presented here are only weakly affected by distance uncertainties,
assumptions about the distribution function in the stellar disk, and
the details of the Galactic potential including the effect of spiral
structure. \Gaia\ and future spectroscopic surveys of the Galactic
disk such as \apogee\ and \hermes\ will be able to robustly test the
origin of the Hercules stream and constrain the properties of the
Galactic bar.
\end{abstract}


\section{Introduction}

\begin{figure}
\includegraphics[width=0.5\textwidth]{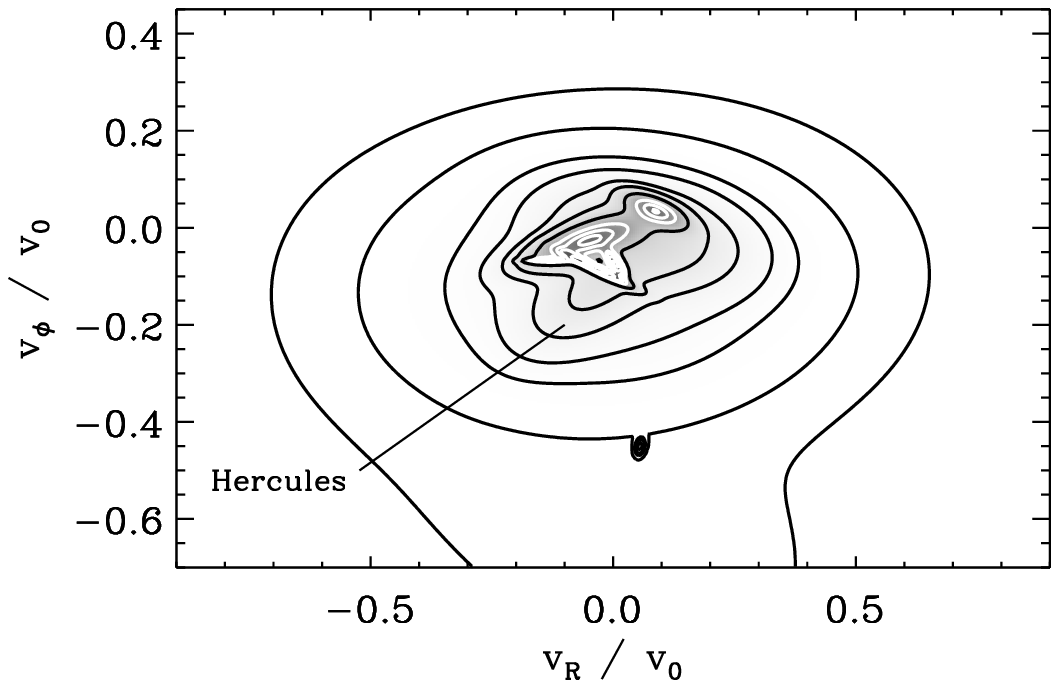}
\caption{Velocity distribution of nearby ($\lesssim 100$ pc) stars
  from \citet{Bovy09a}. \vR\ and \vphi\ are the velocities toward the
  Galactic center and in the direction of Galactic rotation,
  respectively, corrected for the Solar motion. The Hercules moving
  group is visible as an overdensity at \vphi $\approx$ -0.2
  $\vo$.}\label{fig:obs}
\end{figure}

The velocity distribution of nearby stars contains various large
overdensities of co-moving stars that cannot be explained as the
vestiges of clumpy star formation \citep[\eg,][]{Bovy10a}. In
particular, the Hercules moving group, a group of stars significantly
offset from the bulk of the observed velocity distribution
\citep[\eg,][see \figurename~\ref{fig:obs}]{Dehnen98b,Bovy09a},
displays a wide range of ages and metallicities
\citep{Blaauw70a,raboud98a,caloi99a,Bensby07a,Bovy10a}. A compelling
model has the Hercules stream originating through resonant
interactions of stars in the outer disk with the bar in the central
region of the Galaxy \citep{dehnen00a,fux01a}, but there is no
confirmation of this picture beyond the locally observed stellar
kinematics that it was proposed to explain. An investigation of the
Hercules members' color--magnitude diagram found no evidence for any
significant metallicity anomaly with respect to other local stars
\citep{Bovy10a}, even though such a difference should be expected as
the Hercules stars originate from a few kiloparsec toward the Galactic
center where the average metallicity is higher than in the Solar
neighborhood. The absence of the predicted metallicity anomaly could
potentially be explained, \eg, through radial mixing
\citep{sellwood02a}, but it is clear that chemodynamical modeling of
the Galactic disk is not currently up to the task of separating out
these effects \citep[\eg,][]{samland03a,roskar08a}.

Here I propose that a robust and unambiguous test of the bar-resonance
model for the Hercules stream is a search for the distinct pattern it
predicts as we trace the velocity distribution of stars near the Solar
radius around the Galaxy. \Gaia\ is in a good position to have the
final word on this, but I also show that the Hercules feature can be
detected in the line-of-sight velocity distribution in selected
directions on the sky. Clear signatures are predicted to be detectable
in regions a few kiloparsec from the Sun, allowing for a clean
verification of the bar-resonance model for the Hercules stream.

\section{Methodology}\label{sec:method}

To simulate the effect of the bar on stellar orbits near the Solar
circle, we follow the approach of \citet{dehnen00a}. For each fixed
position in the Galaxy, we evaluate the velocity distribution function
at values for the radial---toward the Galactic center---and
tangential---in the direction of Galactic rotation---velocity by
backward-integrating these velocities in the Galactic potential to
obtain the intial orbit before bar formation. We posit that at this
epoch the old stellar disk was in a steady-state that can be described
by a simple distribution function that is a function of the energy and
angular momentum of the orbit alone. It follows from the collisionless
Boltzmann equation that the value of the velocity distribution at the
present position and velocity is equal to that of the steady-state
distribution function before bar formation evaluated at the initial
position and velocity---or, equivalently, at the initial orbit since
the initial distribution function is time-independent.

Since the Hercules stream is only apparent in the planar motions in
the disk \citep[\eg,][]{Bovy09a}, we only consider a two-dimensional
model for the Galaxy. We use a simple power-law rotation curve to
model the axisymmetric Galactic potential
\begin{equation}
v_c(R) = \vo(R/\Ro)^\beta \,,
\end{equation}
where \Ro\ is the distance from the Sun to the Galactic center. For
most of the simulations below we specify this model further to a flat
rotation curve ($\beta = 0$). The model for the bar is given by the
following potential
\begin{equation}
\Phi_b(R,\phi) = \Ab(t) \cos [2(\phi - \Omegab t)] \times
\left\{ \begin{array}{ll} -(\Rb/R)^3\,, & \mathrm{for}\ R \geq
  \Rb\,,\\ (R/\Rb)^3-2\,, & \mathrm{for}\ R \leq \Rb\,. \end{array}
\right.
\end{equation}
Here \Omegab\ is the pattern speed of the bar and \Rb\ is the bar
radius, which is fixed to be 80\,percent of the bar's corotation
radius. The bar is grown smoothly during a time $t_1$, always set to
half of the total integration time in what follows, using the
prescription
\begin{equation}
\Ab(t) = \alpha \, \frac{\vo^2}{3}\,\left(\frac{\Ro}{\Rb}\right)^3
\times \left\{ \begin{array}{lll}
\frac{3}{16}\xi^5 - \frac{5}{8} \xi^3 + \frac{15}{16} \xi + \frac{1}{2}\,, & \xi \equiv 2\frac{t}{t_1}-1\,, & t \leq t_1\,,\\
1\,, & & t \geq t_1\,.\end{array} \right. 
\end{equation}
A comparison of the local velocity distribution obtained by using this
simple bar potential with that resulting from the more realistic one
of \citet{gardner10a} shows that this simple potential is sufficient
to model the bar's influence on stellar orbits in the outer disk.

For the distribution function of the stellar disk before bar
formation, we use a Dehnen distribution function \citep{dehnen99b}
given by
\begin{equation}\label{eq:fdehnen}
\fdehnen(E,L) \propto \frac{\Sigma(\rE)}{\sigmaR^2(\rE)} \, \exp\left[ \frac{\Omega(\rE)\left[L-\Lc(E)\right]}{\sigmaR^2(\rE)}\right]\,,
\end{equation}
where \rE, \Lc, and $\Omega(\rE)$ are the radius, angular momentum,
and angular frequency, respectively, of the circular orbit with energy
$E$. Using the procedure given in \sectionname~3.2 of
\citet{dehnen99b}, we choose the $\Sigma(R)$ and $\sigmaR(R)$
functions such that they reproduce a disk with exponential surface
density and velocity dispersion profiles
\begin{equation}
\Sigma(R) = \Sigma_0 \exp\left(-R/\Rs\right)\,, \qquad 
\sigmaR(R) = \sigma_0 \exp\left(-R/\Rsigma\right)\,,
\end{equation}
to an accuracy of a fraction of a percent at all radii. Our fiducial
model for the distribution function has \Rs = \Ro /3 and \Rsigma =
3\Rs, and $\sigma_0$ such that $\sigmaR(\Ro) = 0.2 \vo$.

As shown by \citet{dehnen99c,dehnen00a}, the model described above
with the Sun at \Ro = 0.9\Rolr---where \Rolr\ is the radius at which a
circular orbit is in the outer Lindblad resonance with the bar---at a
current angle of 25$^{\circ}$ with the bar, with a bar strength
$\alpha = 0.01$, and integrating for four bar periods, reproduces the
Hercules feature in the local velocity distribution while agreeing
with other photometric and kinematical observations of the bar
\citep[\eg,][]{binney97a,bissantz02a,cole02a,Shen10a}. In what
follows, unless indicated otherwise, we fix all of the parameters of
the model at these fiducial values. All velocities are always
considered to be with respect to their local standard of rest in the
absence of a bar.

\section{The full planar velocity distribution in the disk}\label{sec:2d}

\citet{dehnen00a} only simulated the velocity distribution at the
position of the Sun for various combinations of the model parameters
defined in the previous Section. Nevertheless, his results on how the
velocity distribution at the Sun changes with the angle between the
Sun-Galactic center line and the bar could also be interpreted as
showing the velocity distribution on the Solar circle for different
Galactocentric azimuths, fixing the bar angle. In this way, these
simulations can make predictions for what we might find when we study
the velocity distribution in regions beyond those probed by the
\hipparcos\ mission \citep{ESA97a}.

\begin{figure}
\includegraphics[width=\textwidth]{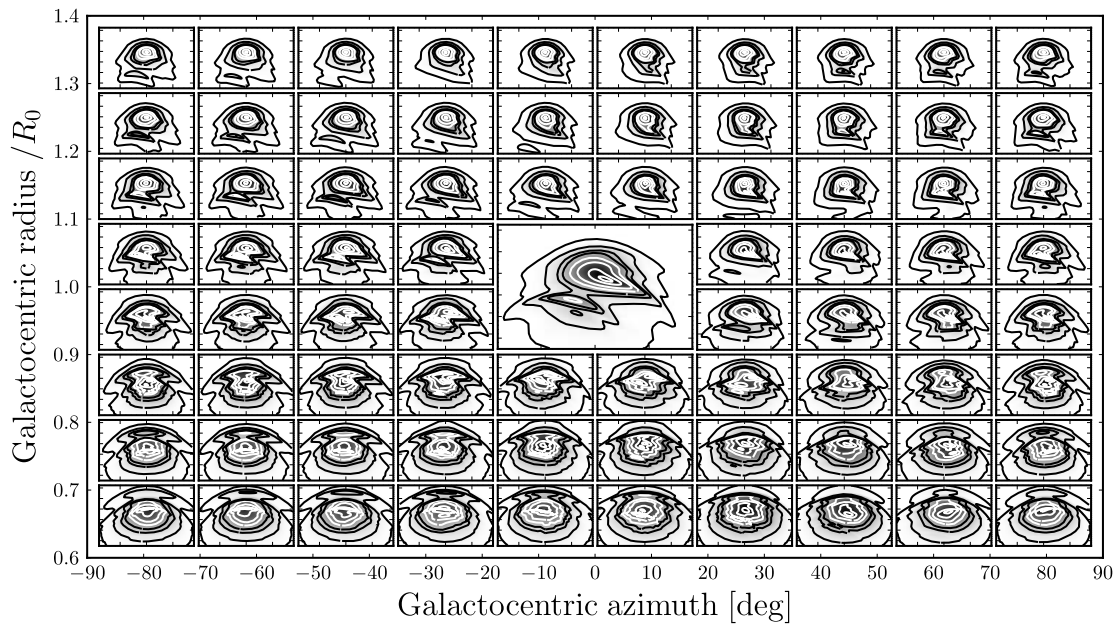}
\caption{Two-dimensional velocity distributions along the Solar circle
in the bar-resonance model for the Hercules stream. The $x$-axis in
each subplot is the Galactocentric radial velocity and the $y$-axis is
the Galactocentric tangential velocity; the axis-ranges of each
subplot are the same as in \figurename~\ref{fig:obs}. The velocity
distributions on the other side of the Galaxy are obtained by shifting
the azimuths by 180$^{\circ}$.}\label{fig:rphi2d}
\end{figure}

In \figurename~\ref{fig:rphi2d}, I present the predictions for the
two-dimensional velocity distributions near the Solar circle for the
full 360$^{\circ}$ range in Galactocentric azimuth for the fiducial
model from \sectionname~\ref{sec:method}. Only the range $-90^{\circ}
\leq \phi \leq 90^{\circ}$ is shown here, since the barred Galaxy is
symmetric with respect to rotations through 180$^{\circ}$; the
predictions for the other side of the Galaxy can be obtained from
those shown by shifting the azimuths by 180$^{\circ}$. The predicted
velocity distributions on the Solar circle are not shown here, but
they can be inferred from \figurename~2 in \citet{dehnen00a}.

It is clear from this \figurename\ that we can trace the Hercules
stream going around the Galaxy and when looking further in toward the
Galactic center or further out. Furthermore, the effect of the bar is
not confined to a narrow range in radii, but rather the effects of the
bar and especially its outer Lindblad resonance are felt through a
large range of radii (0.6 $\leq R/\Ro \leq 1.4$). These effects are
strong and robust predictions of the bar-origin model for the Hercules
stream. This is especially the case in regions beyond the Solar
radius. There the Hercules feature can be found significantly removed
from the bulk of the velocity distribution, which might be severely
shaped at all locations by the processes that create the pattern of
low-velocity moving groups in the local velocity distribution (the
main mode in \figurename~\ref{fig:obs}). As I discuss below, \Gaia\
will be able to verify these predictions.

\section{Line-of-sight velocity distributions}

Rather than wait for a comprehensive survey of the Galactic disk that
measures the full phase-space position of a large number of stars by
measuring proper motions, we can ask whether a targeted survey of
line-of-sight velocities in selected regions of the Galaxy could
detect a clear signature of the Hercules feature? The Hercules feature
stands out in the velocity distribution most in its tangential
velocity. On the Solar circle near the Sun the line-of-sight velocity
is almost entirely this Galactocentric tangential velocity, so we
would expect to see at least a hint of the Hercules stream in the
line-of sight.

\begin{figure}
\includegraphics[width=\textwidth]{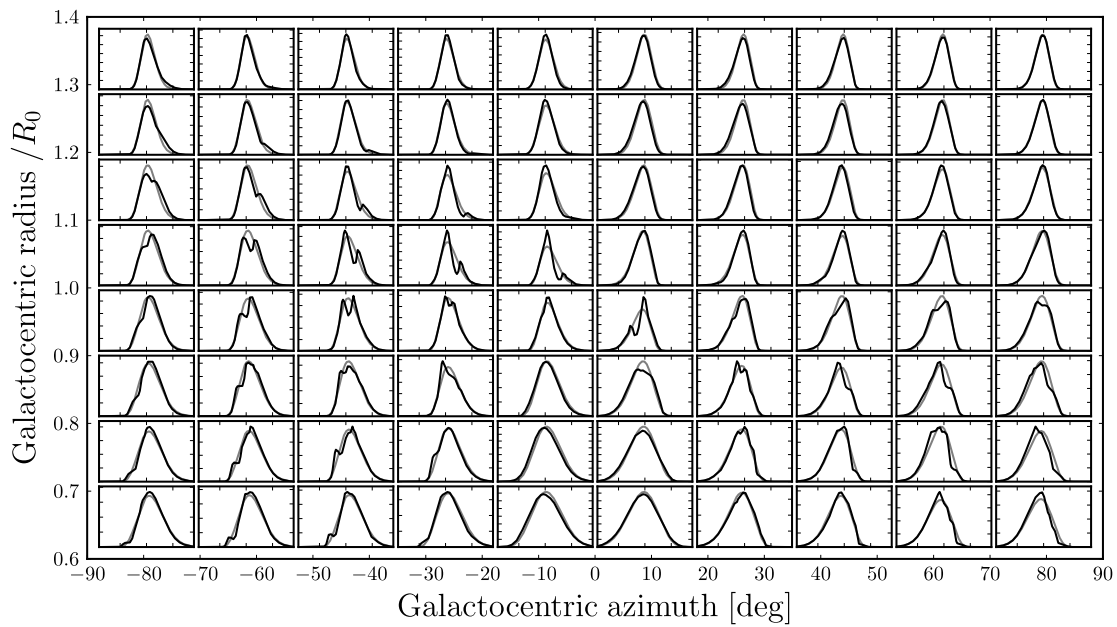}\\
\includegraphics[width=\textwidth]{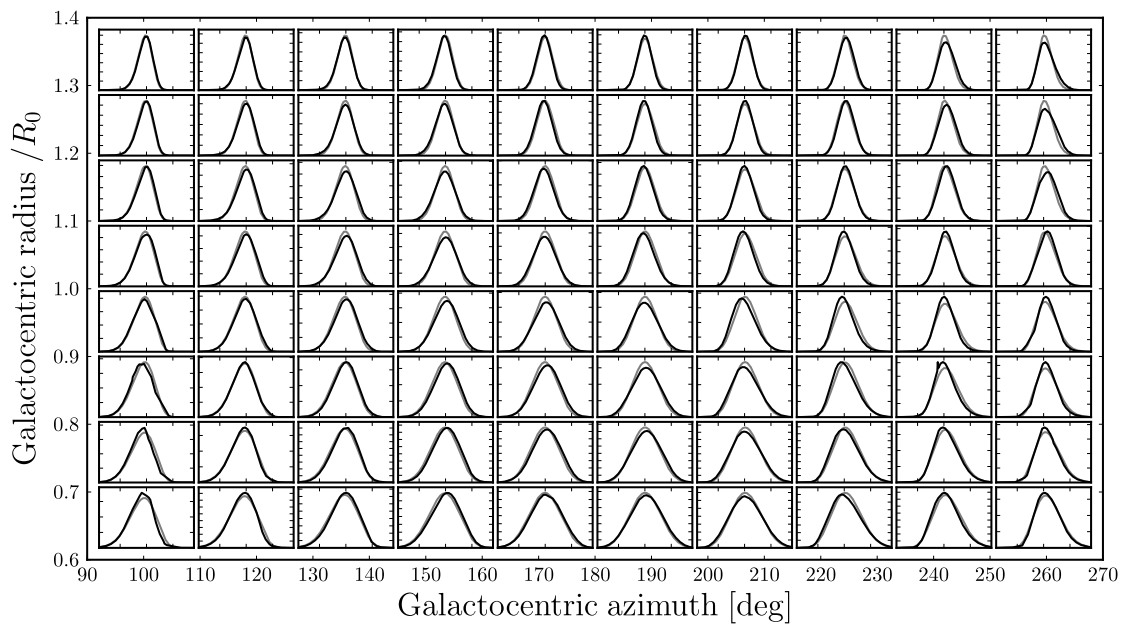}
\caption{Line-of-sight velocity distributions along the Solar
  circle. The range in line-of-sight velocities in each subpanel is as
  in \figurename~\ref{fig:1dvar}. The gray curve in each panel is the
  predicted distribution for an axisymmetric, steady-state
  Galaxy.}\label{fig:rphi1d}
\end{figure}

In \figurename~\ref{fig:rphi1d} I show the predicted line-of-sight
velocity distributions in various locations near the Solar circle all
around the Galaxy, obtained by integrating the full two-dimensional
velocity distribution over the velocity component perpendicular to the
line of sight. This \figurename\ is no longer symmetric with respect
to shifts of 180$^{\circ}$ in azimuth, since the line-of-sight
direction breaks this symmetry. The gray curve is the line-of-sight
distribution we would observe if the targeted stars were in a steady
state in an axisymmetric potential with a Dehnen distribution function
(\eqnname~[\ref{eq:fdehnen}]). Both line-of-sight velocity
distributions are normalized to have equal mass since I assume that
our knowledge of the Galaxy and the tracer population will not be good
enough that we can test predicted differences in the absolute number
of stars at a given location between the axisymmetric and the
bar-resonance predictions. These differences are also largely a
function of the particular form and parameters of the distribution
function that we choose to represent the stellar disk before bar
formation.

The prospects to detect the Hercules feature are quite good on our
side of the Galaxy. The Hercules feature is especially strong in the
line-of-sight velocity distribution for Galactic longitudes
$250^{\circ} \lesssim l \lesssim 290^{\circ}$. This is because of the
fortunate coincidence that for those lines-of-sight we see velocities
in the direction along which the Hercules streams separates most from
the bulk of the velocity distribution (see
\figurename~\ref{fig:rphi2d}) and that at these azimuths the Hercules
stream is at the furthest distance from the rest of the velocity
distribution. We see hardly any effect of the bar on the line-of-sight
velocity distributions on the other side of the Galaxy ($90^{\circ}
\leq \phi \leq 270^{\circ}$). On that side we mostly see the
Galactocentric radial velocity in the line-of-sight, in which the
Hercules feature does not separate clearly from the bulk of the
stars. Even though upcoming surveys such as \apogee\ have the ability
to probe the other side of the Galaxy, for the purposes of detecting
the Hercules stream this is uninteresting as it does not show up in
the line-of-sight velocity distribution there.

To further analyze the directions to target to best see the Hercules
feature in the line-of-sight velocities, I have calculated the
line-of-sight velocity distribution on a 100 by 20 grid in
Galactocentric azimuth and radius. As a measure of the difference
between the bar-resonance model and the steady-state, axisymmetric
model, I have calculated the Kullback-Leibler divergence between these
predicted distributions \citep[\eg,][]{mackay}
\begin{equation}
D_{\text{KL}}(P||Q) = \int \dd x\,p(x)\,\ln \frac{p(x)}{q(x)}\,,
\end{equation}
where, in this case, $x$ is the line-of-sight velocity, $p(x)$ is the
predicted distribution for the bar-origin model of the Hercules
stream, and $q(x)$ is the axisymmetric prediction. The
Kullback-Leibler divergence is only one among many possible choices to
quantify the difference between two probability distributions. It can
be interpreted as the expected weight of evidence for the prediction
by the bar-resonance model over the fully axisymmetric prediction that
can be expected from each sampled star. As such, locations with the
maximal Kullback-Leibler divergence are those in which the Hercules
feature stands out most prominently in the line of sight.

The result of this analysis is shown in
\figurename~\ref{fig:detect}. I present results both using the full
line-of-sight velocity distributions and only considering the part of
the distribution at $|v_{\text{los}}| \geq 0.15 \vo$. As is clear from
\figurename~\ref{fig:obs}, the local velocity distribution contains
many more features than the Hercules stream. These features are mostly
contained to the main mode of the velocity distribution $|v| \lesssim
0.15 \vo$. Observations indicate that these features are also the
result of dynamical interactions \citep[\eg,][]{Bovy10a,sellwood10a},
albeit with spiral arms rather than the bar. Therefore, we can expect
these or similar features to be present in the velocity distribution
at other locations in the Galaxy as well, and thus, if the effect of
the Hercules stream occurs close to the center of the bulk of the
velocity distribution, it might be hard to disentangle from the other
dynamical effects at play. However, these additional dynamical effects
will be much less important for velocities $\gtrsim 0.15
\vo$. \figurename~\ref{fig:detect} shows that there are many
locations, mostly at $250^{\circ} \lesssim l \lesssim 290^{\circ}$,
where the Hercules stream is clearly offset from the bulk of the
velocity distribution and should be unambiguously detectable.

Finally, in \figurename~\ref{fig:1dvar} I show the influence of
various nuisance and dynamical parameters on the predicted
line-of-sight velocity distribution for the most promising region that
is more than 3 kpc away. We see in this \figurename\ that the effect
of distance and realistic line-of-sight velocity uncertainties is
negligible. The exact shape and parameters of the stellar distribution
function also only minimally affect the predicted line-of-sight
velocity distribution. Changing the scale length within its current
uncertainty changes the prediction only in the amplitude of the
effect: The Hercules stream is more prominent for shorter scale
lengths since then there are relatively more stars in the inner Galaxy
that are brought to the Solar circle by the interaction with the
bar. Using a Shu distribution function \citep{shu69a} instead of the
Dehnen distribution function also gives very similar
predictions. Changing the axisymmetric potential by adjusting the
shape of the rotation curve shifts the peak as expected
\citep[see][]{dehnen00a}, but the Hercules feature nevertheless
remains detectable. The properties of the bar itself, \eg, its pattern
speed through \Rolr\ or its strength $\alpha$, change the predicted
pattern, but they do not significantly alter the prospects for
detection.

\section{Discussion}

The discussion in the previous sections shows that the bar-origin
paradigm for the Hercules moving group makes strong predictions about
how the Hercules feature should show up when we probe the velocity
distribution in parts of the Galaxy far removed from the Solar
neighborhood. In order to most strongly test the bar origin for the
Hercules moving group it is best to extrapolate this picture far from
the Solar neighborhood and look at regions further than a few kpc
removed. \Gaia 's parallaxes, proper motions, and mission-averaged
line-of-sight velocities for dwarfs nearby and giants at distances
greater than a few kpc will allow the full two-dimensional velocity
distribution to be confronted with the predictions of
\figurename~\ref{fig:rphi2d} (see below). However, the final
\Gaia\ catalog will only be available around 2020.

The predicted line-of-sight velocity distributions in the previous
section show that the bar-origin theory for the Hercules moving group
can be tested with line-of-sight velocities alone. At distances
greater than a few kpc this will mean observing giants to trace the
velocity distribution. Large, planned spectroscopic surveys with
line-of-sight velocity uncertainties less than 1 km s$^{-1}$ can test
the predictions made in this paper in the next few years (see
below). These surveys will not need to observe many stars to test the
predictions laid out here: Drawing samples from the predicted velocity
distribution in \figurename~\ref{fig:1dvar} shows that about 500 stars
suffice to distinguish the bar-origin model from the axisymmetric
prediction (based on a KS test with $P < 0.01$). Since large distance
uncertainties do not affect the prediction very much (see
\figurename~\ref{fig:1dvar}), these stars can even be sampled
throughout a volume of 500 pc or more. These spectroscopic surveys
will also obtain detailed elemental abundances for their target stars,
so they would also allow for a more detailed analysis of the
abundances of stars in the Hercules stream \citep[following,
  \eg,][]{Bensby07a,Bovy10a}, which could shed more light on radial
mixing in the Galactic disk.

The last two panels of \figurename~\ref{fig:1dvar} show that the
predicted line-of-sight velocity distribution depends strongly on the
dynamical parameters of the bar. Changing the pattern speed of the bar
through \Rolr\ shifts the predicted distribution and in particular the
location of the Hercules feature. While this shift at a particular
location is degenerate with the shift due to the uncertainty in the
local circular velocity, results from various locations can be
combined to break this degeneracy and the local circular velocity is
also already quite well constrained \citep[\eg,][]{Bovy09b}. Changing
the strength of the bar changes the relative heights of the two main
peaks in the predicted velocity distribution. However, the main peak
of the velocity distribution might not be as simple as the one
predicted here due to other dynamical effects. Integrated measures,
such as the number of stars in the main peak and the number of stars
in the Hercules feature, could mitigate this somewhat. While
predicting actual constraints would involve detailed simulations of
the expected data, it is clear that constraints on the dynamical
properties of the bar can be derived from the experiment proposed in
this paper.

\subsection{\Gaia}

The astrometric \Gaia\ mission will measure the parallaxes and proper
motions of up to one billion stars, most of which will be disk stars
\citep{Perryman01a}. With parallax and proper motion accuracies down
to 10 $\mu$as and mission-averaged line-of-sight velocity
uncertainties smaller than 10 km s$^{-1}$ for many stars to 17th
magnitude, \Gaia\ will be able to probe the kinematics of the disk out
to several kpc in all directions
\citep{Katz04a,bailerjones08a}. Crucially, it will provide large
samples of giants in many of the promising regions in
\figurename~\ref{fig:rphi2d}, with velocity measurements accurate
enough to verify the predictions laid out in this paper. This will
also allow the dynamical properties of the bar to be tightly
constrained, something that will be hard for \Gaia\ to do using direct
observations of the bulge due to the large extinction toward the
Galactic center \citep{robin05a}.

\subsection{Radial velocity surveys}

\begin{figure}
\includegraphics[width=0.5\textwidth]{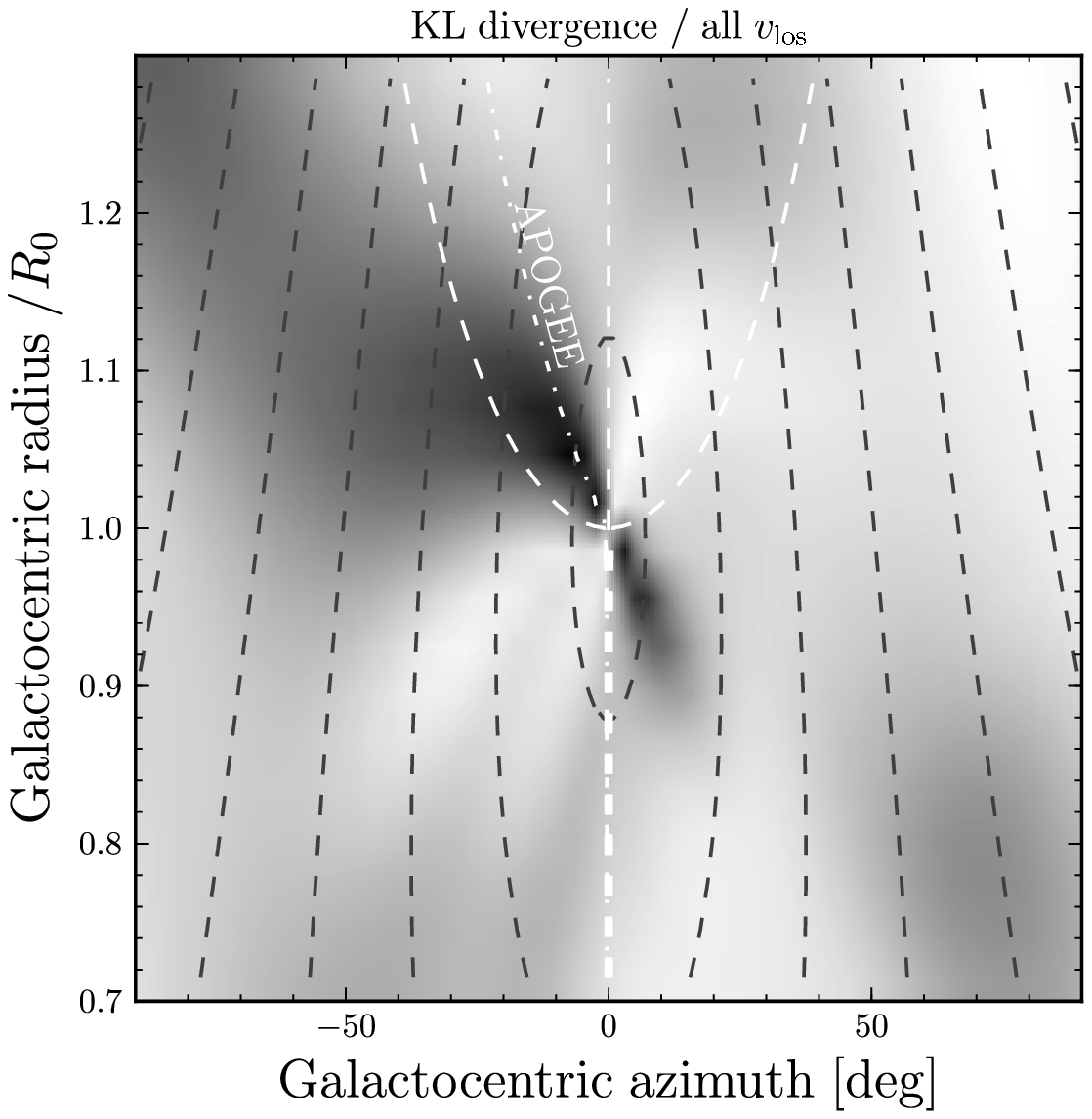}
\includegraphics[width=0.5\textwidth]{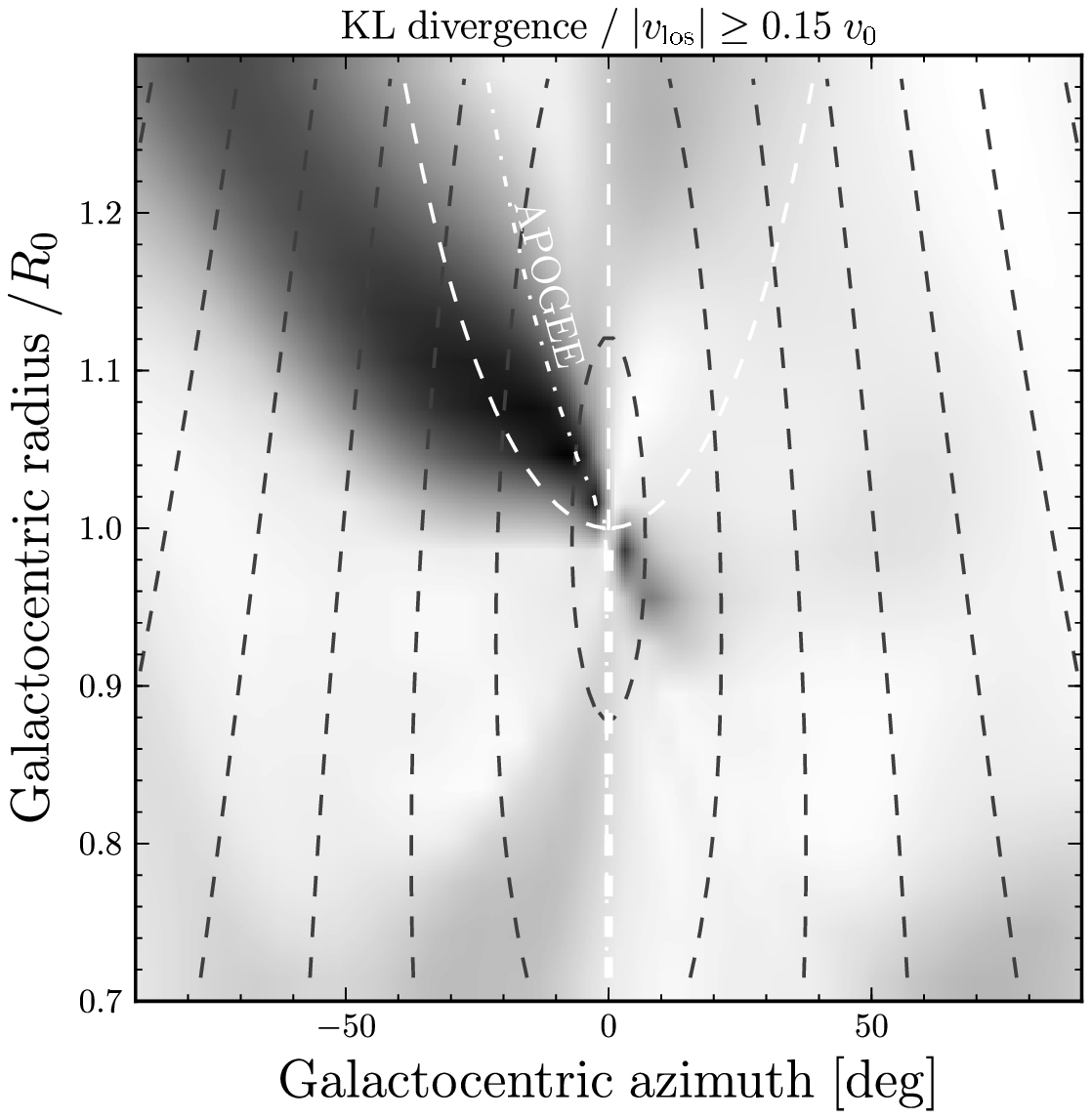}
\caption{Kullback-Leibler divergence between the predicted
  line-of-sight velocity distributions with and without the
  bar. Darker values have a larger discriminatory potential. The left
  panel uses the full line-of-sight velocity distributions, the right
  panel excludes the main mode of the velocity distribution that is
  likely to be affected by other dynamical and phase-space
  effects. The concentric dashed lines indicate constant-distance
  slices from the Sun, starting at 1 kpc and spaced 2 kpc apart.  The
  white dashed lines indicate the boundaries of the four quadrants in
  Galactic longitude.  The white dashed-dotted line shows the region
  of the sky that \apogee\ can observe (-5$^{\circ} \leq l \leq
  250^{\circ}$).}\label{fig:detect}
\end{figure}

\begin{figure}
\includegraphics[width=0.32\textwidth]{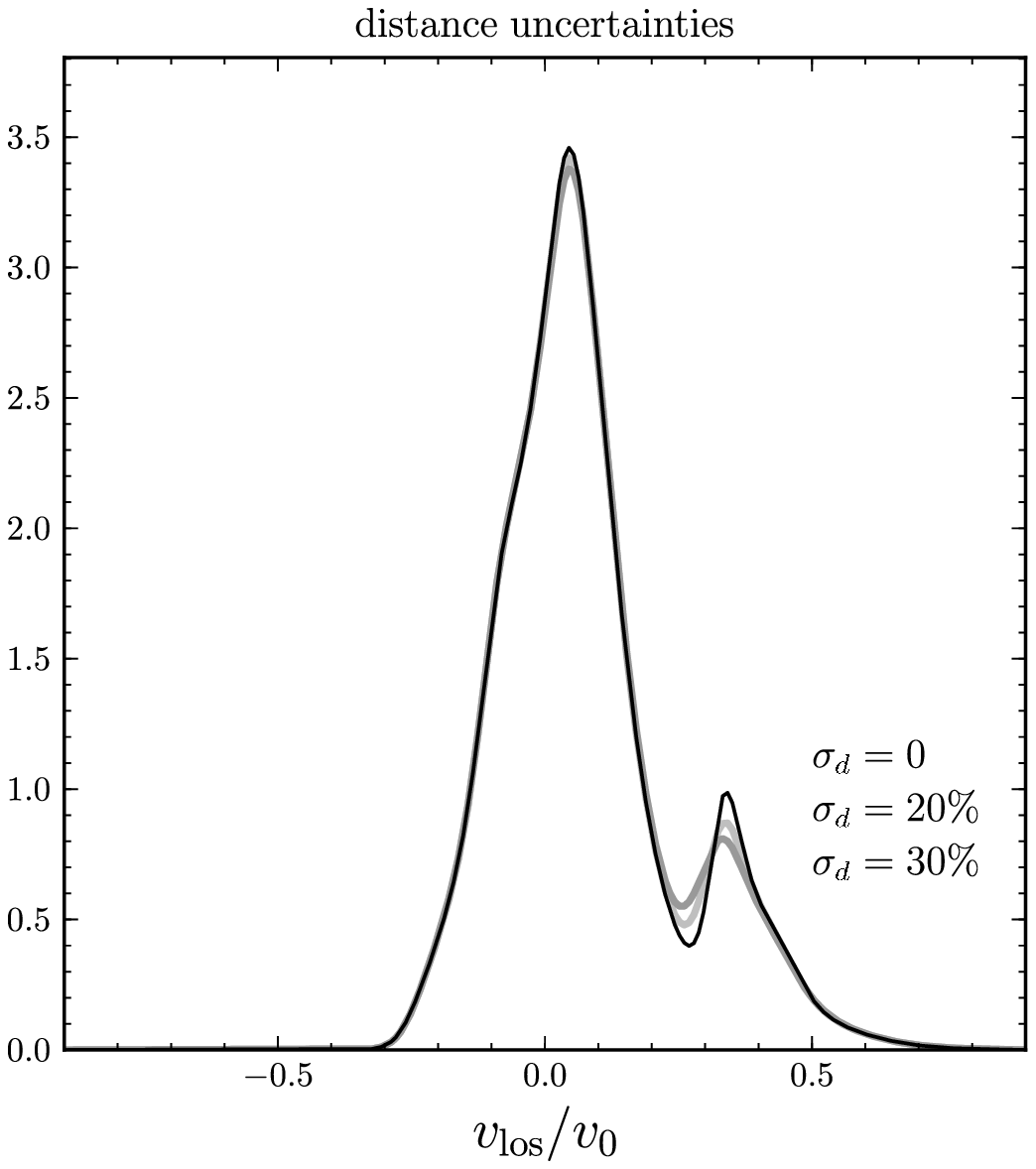}
\includegraphics[width=0.32\textwidth]{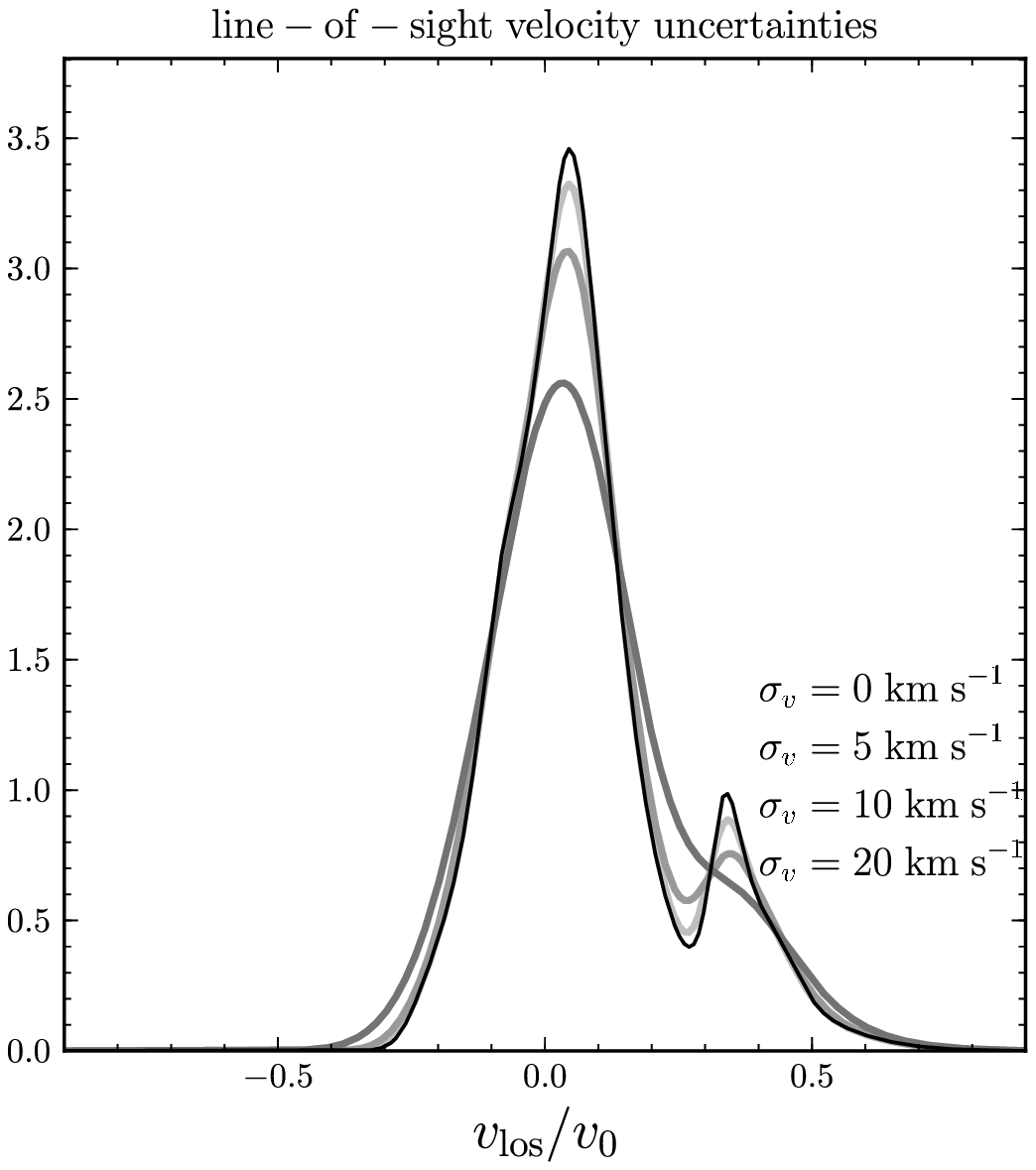}
\includegraphics[width=0.32\textwidth]{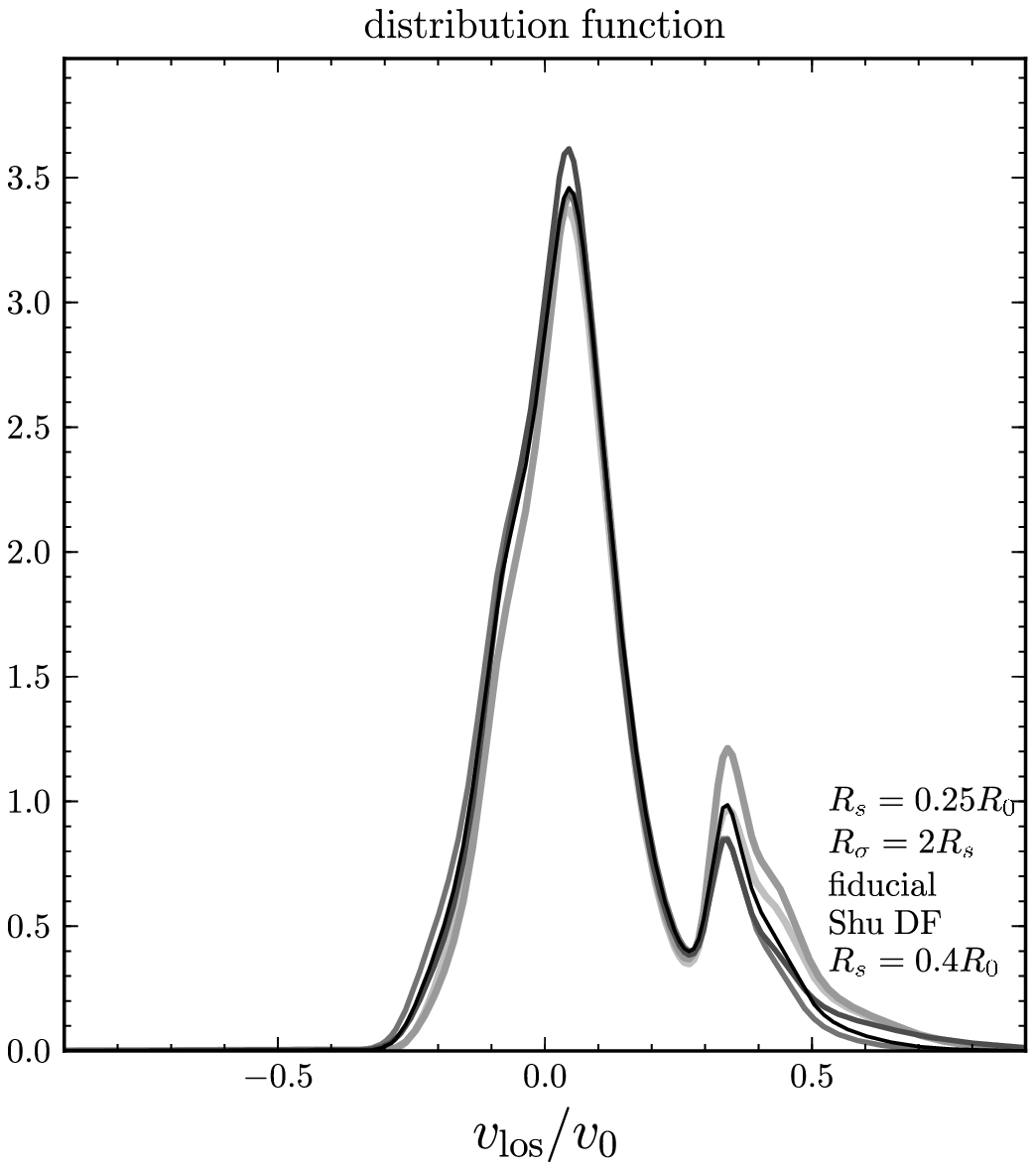}\\
\includegraphics[width=0.32\textwidth]{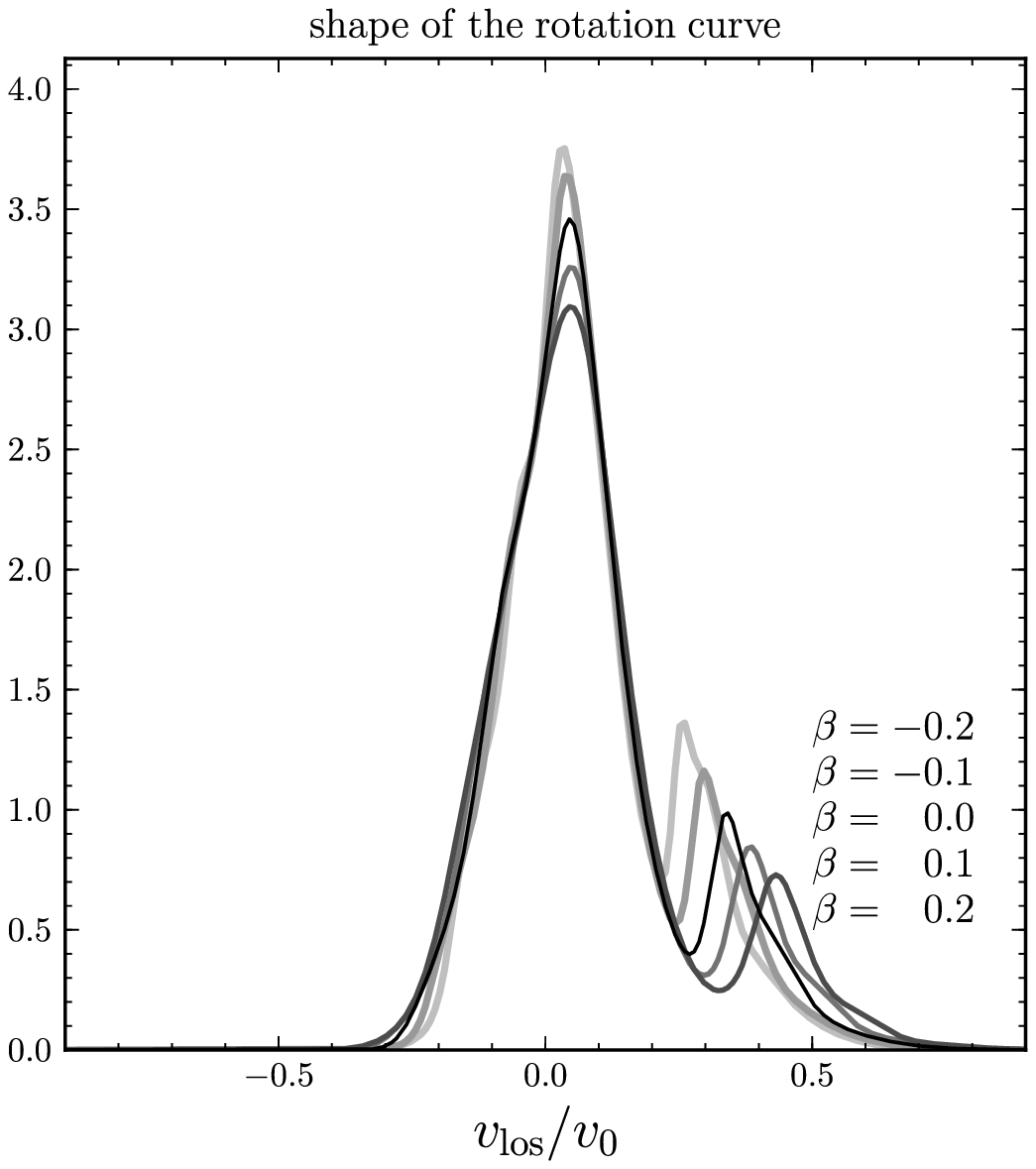}
\includegraphics[width=0.32\textwidth]{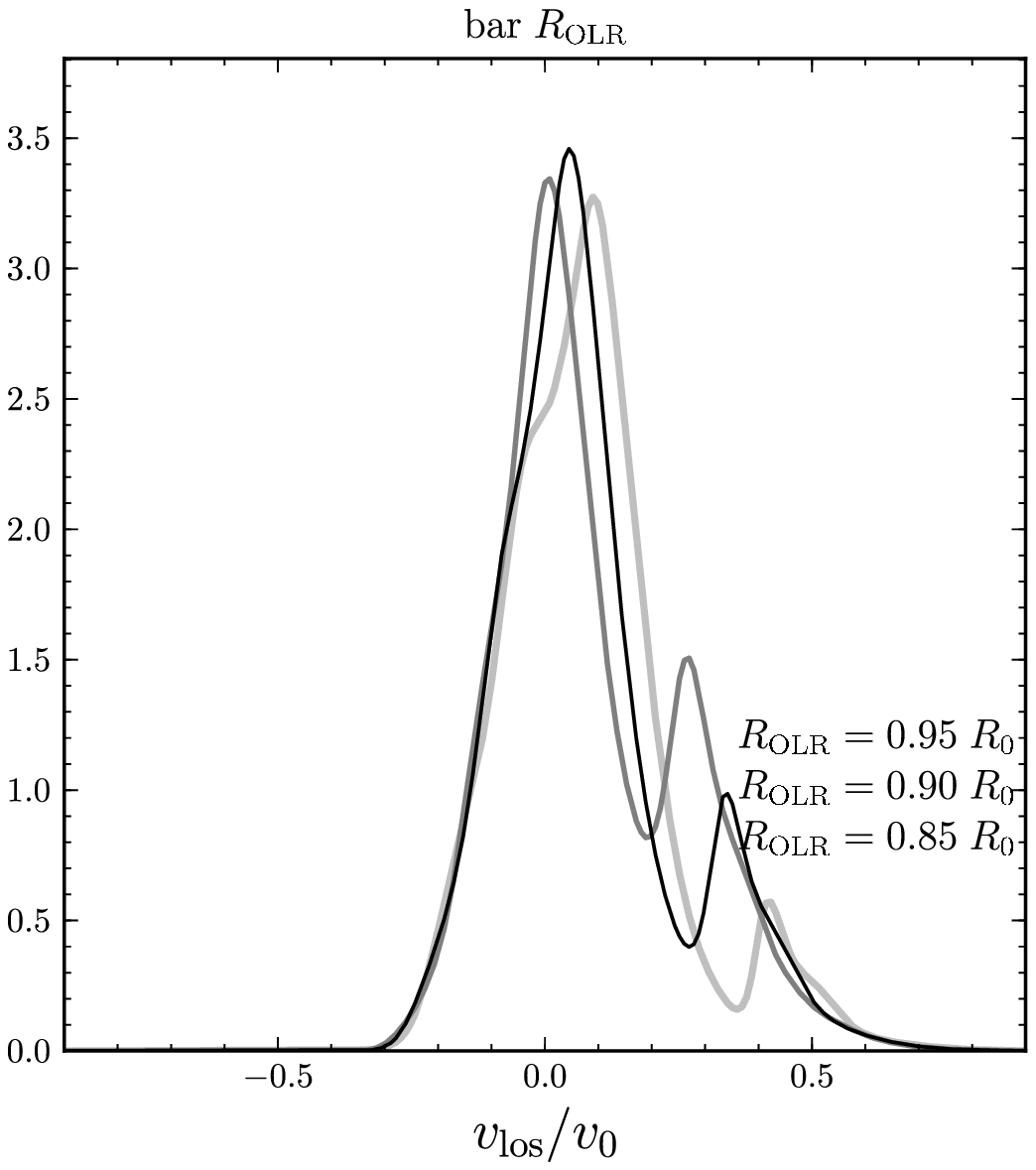}
\includegraphics[width=0.32\textwidth]{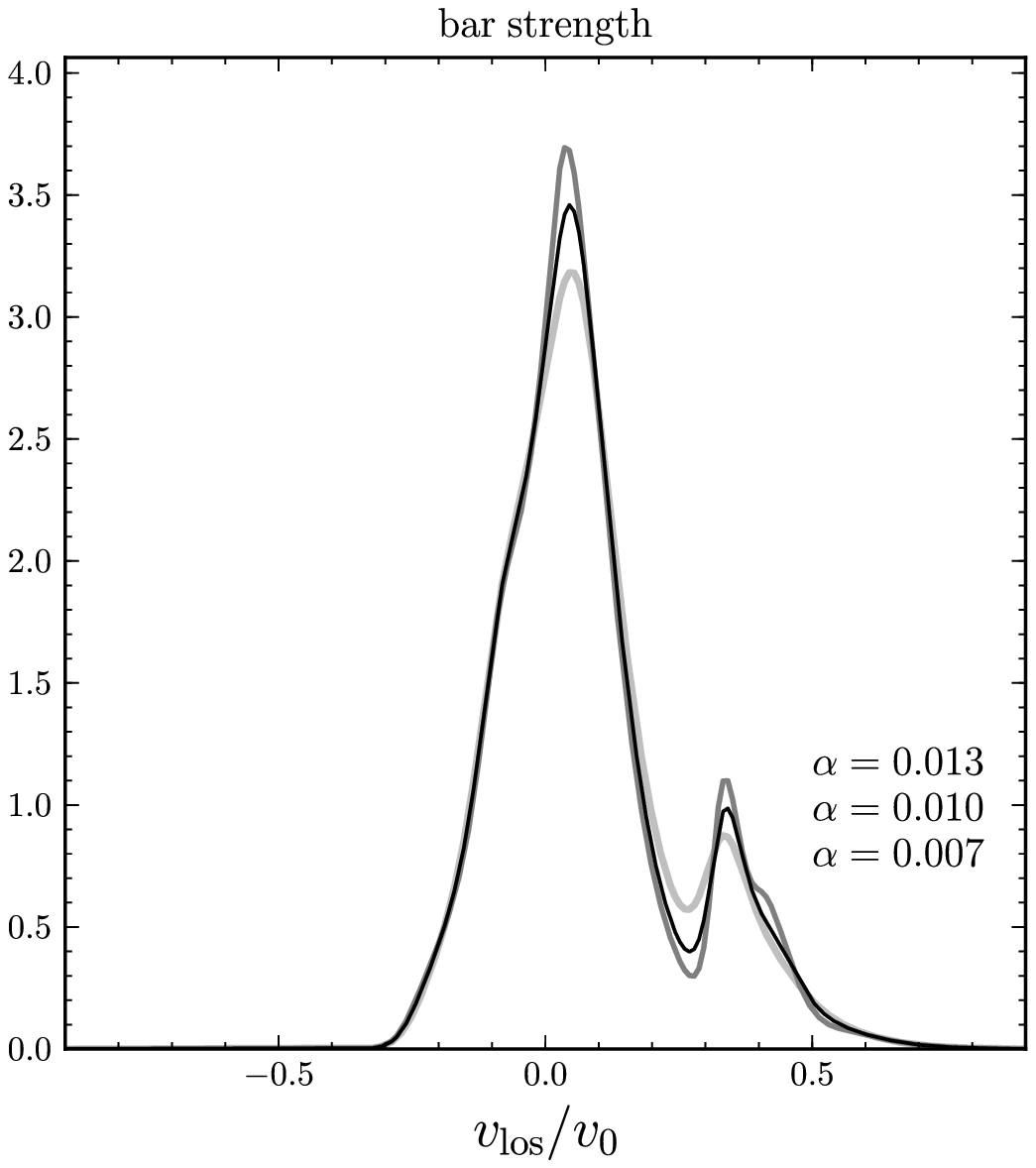}
\caption{Variation of the predicted velocity distribution at $R$ =
  1.075 \Ro , $\phi = -21^{\circ}$ ($d$= 0.4 \Ro, $l$ = 270$^{\circ}$)
  with distance uncertainties; line-of-sight velocity uncertainties;
  parameters and shape of the distribution function; shape of the
  rotation curve; bar \Rolr; and bar strength.}\label{fig:1dvar}
\end{figure}

Since the extinction in the optical bandpasses is rather large in the
directions identified as promising in \figurename~\ref{fig:detect}
\citep[$A_V \approx 8$ mag toward $l \approx
  270^{\circ}$;][]{marshall06a}, spectroscopic surveys in the
near-infrared such as \apogee\ are the best option for seeing the
Hercules feature in the line-of-sight direction. The range of Galactic
longitudes that \apogee\ can target along the disk is $-5^{\circ} \leq
l \leq 250^{\circ}$ and its expected line-of-sight uncertainties are
about 0.5 km s$^{-1}$ (SDSS-III Project Description\footnote{Available
  at \url{http://sdss3.org/collaboration/description.pdf}~.}). From
\figurename~\ref{fig:detect} this longitude limit is right on the edge
of where we might hope to detect the Hercules feature. The Hercules
bump in the direction $l = 250^{\circ}$ is about half as strong as in
\figurename~\ref{fig:1dvar}, but remains as a clear signature of the
bar. As before, about 500 stars a few kpc away are enough to detect
the Hercules feature. Since \apogee\ is expected to start taking data
in 2011 and to finish operations before the first \Gaia\ data release,
\apogee\ could potentially trace the Hercules stream around the Galaxy
before \Gaia\ does.

Another planned spectroscopic survey is
\hermes\ \citep{Freeman10a}. While operating in the optical, it
expects to survey hundreds of thousands of giants in the inner Galaxy,
many as far as 6 kpc away, with line-of-sight velocity uncertainties
less than 1 km s$^{-1}$ (K. Freeman, private communication). Since
\hermes\ operates from the Southern hemisphere it can observe the most
promising directions in \figurename~\ref{fig:detect} and, like
\apogee, detect the Hercules feature before \Gaia.

\acknowledgements It is a pleasure to thank the anonymous referee for
valuable comments and Mike Blanton and David W.~Hogg for helpful
discussions and assistance.  Financial support for this project was
provided by the National Aeronautics and Space Administration (grant
NNX08AJ48G) and the National Science Foundation (grant AST-0908357).

\clearpage

\clearpage

\clearpage

\clearpage

\clearpage

\end{document}